# Nitrogen-Based Magneto-Ionic Manipulation of Exchange Bias in CoFe/MnN Heterostructures

Christopher J. Jensen[1], Alberto Quintana[1], Patrick Quarterman[2], Alexander J. Grutter[2], Purnima P. Balakrishnan[2], Huairuo Zhang[3,4], Albert V. Davydov[4], Xixiang Zhang,[5] and Kai Liu[1,*]

[1]Physics Department, Georgetown University, Washington, DC 20057, USA
[2]NIST Center for Neutron Research, NCNR, National Institute of Standards and Technology, Gaithersburg, MD 20899, USA
[3]Theiss Research, Inc. La Jolla, California 92037, USA
[4]NIST Materials Measurement Laboratory, National Institute of Standards and Technology, Gaithersburg, MD 20899, USA
[5]King Abdullah University of Science & Technology, Thuwal 23955-6900, Saudi Arabia

**Abstract**

Electric field control of the exchange bias effect across ferromagnet/antiferromagnet (FM/AF) interfaces has offered exciting potentials for low-energy-dissipation spintronics. In particular, the solid state magneto-ionic means is highly appealing as it may allow reconfigurable electronics by transforming the all-important FM/AF interfaces through ionic migration. In this work, we demonstrate an approach that combines the chemically induced magneto-ionic effect with the electric field driving of nitrogen in the Ta/Co$_{0.7}$Fe$_{0.3}$/MnN/Ta structure to electrically manipulate exchange bias. Upon field-cooling the heterostructure, ionic diffusion of nitrogen from MnN into the Ta layers occurs. A significant exchange bias of 618 Oe at 300 K and 1484 Oe at 10 K is observed, which can be further enhanced after a voltage conditioning by 5% and 19%, respectively. This enhancement can be reversed by voltage conditioning with an opposite polarity. Nitrogen migration within the MnN layer and into the Ta capping layer cause the enhancement in exchange bias, which is observed in polarized neutron reflectometry studies. These results demonstrate an effective nitrogen-ion based magneto-ionic manipulation of exchange bias in solid-state devices.

Electric field control of the exchange bias (EB) effect across ferromagnet/antiferromagnet (FM/AF) interfaces[1] has offered exciting potentials for low-energy-dissipation spintronics, as it is central to spin-valve type of devices such as magnetic tunnel junctions (MTJs).[2-5] To date, a number of approaches have shown promise in this regard, based on multiferroics,[6-7] solid state magneto-ionics,[8-12] memristors,[13] electrolytes,[14] and spin-orbit torque.[15] Among them, solid state magneto-ionics is particularly appealing as it allows reconfigurable electronics by transforming the all-important FM/AF interfaces through ionic migration and enabling a wide variety of magnetic functionalities, such as magnetic anisotropy,[16-18] antiferromagnetism,[8-10] ferromagnetism,[19-25] ferrimagnetic order,[11] and Dzyaloshinskii-Moriya interaction and spin textures.[26-30] Magneto-ionic (MI) control of EB has so far been demonstrated in several oxide systems.[8-10] For example, in Gd/NiCoO,[9] a FM NiCo layer was created through the spontaneous redox reaction at the interface caused by the Gd affinity to oxygen. After establishing EB between FM NiCo and AF NiCoO, oxygen ions were driven toward Gd through voltage control, resulting in an enhancement in EB.

Alternative ionic species have also been explored for magneto-ionics, such as hydrogen,[11-12, 18, 27-28] nitrogen,[25, 31-32] and hydroxide,[33-34] in the quest to overcome the limitations on room temperature ionic migration and irreversibility seen in certain oxygen-based MI systems. Initial studies in nitrogen-based magneto-ionics have demonstrated faster ionic motion,[25, 31] while also maintaining compatibility with current CMOS technology. Nitrogen diffusion and its impact on exchange bias have also been reported, where chemically-induced diffusion between AF MnN and Ta seed layers have been shown to alter the resultant exchange bias in both in-plane[35-38] and out-of-plane systems,[39] indicating a potential magneto-ionic handle to control exchange bias.

In this work we demonstrate a nitrogen-based magneto-ionic enhancement of exchange bias in $Co_{0.7}Fe_{0.3}$/MnN thin films that can be electrically manipulated. After films are exposed to elevated temperatures during the field cooling process, nitrogen is observed to move out of the MnN layer into both the buffer and capping Ta layers due to the Ta affinity to nitrogen. Under electric field gating, N can be driven back into the MnN layer, leading to a significant enhancement in exchange bias. This effect can be reversed under an opposite gating for a longer duration. Scanning transmission electron microscopy (STEM), X-ray diffraction (XRD), and polarized neutron reflectometry (PNR) provide direct evidence of structural and compositional changes that occur after field cooling and after voltage application. This study thus demonstrates an effective pathway for the magneto-ionic manipulation of exchange bias in solid-state configuration in a nonvolatile and energy-efficient manner.

**Results**

Thin film samples of Ta seed (10 nm)/MnN (30 nm)/$Co_{0.7}Fe_{0.3}$ (1 nm)/Ta (10 nm)/Pd or Pt contact (10nm) (bottom to top structure) were grown by magnetron sputtering on thermally oxidized *p*-type Si substrates ($SiO_2$ = 285 nm). They were field-cooled in a superconducting quantum interference device (SQUID) magnetometer from 700 K, above the Néel temperature of MnN, in a 6.5 kOe [1 Oe = 0.1 mT/$\mu_o$] in-plane magnetic field to 300 K or 10 K. Hysteresis loops were measured for samples in the as-grown (AG), field cooled (FC), and subsequent voltage conditioned (VC) states (FC+VC and FC+VC-VC). Here, FC+VC refers to a positive gating of 20 V for 1 hour after field cooling. The FC+VC-VC refers to the subsequent negative biasing for 1 hour after the initial positive gating. Voltage was chosen to produce a ~0.6 MV/cm electric field, shown to be strong enough to promote changes during sample optimization, and demonstrated to be sufficient in driving ionic migration in other magneto-ionic systems.[9, 17-18] Positive and negative bias refers to having the positive connection on the top contact and the conducting *p*-type Si substrate, respectively. Fig. 1a illustrates the sample structure with the positive gating convention. Further details are provided in Methods.

The AF MnN chosen here is the $\theta$-phase Mn$_6$N$_{5+x}$, where $x \geq 0$, with a face-centered tetragonal (FCT) structure (F4/mmm)[40] and a Néel temperature of $T_N \sim 660$ K.[41-42] The lattice parameters of $\theta$-MnN depend on the N-concentration, with increasing N incorporation leading to larger *a* and *c* lattice constants.[40, 43] Another AF phase, $\eta$-Mn$_3$N$_2$, shares a similar structure with the $\theta$-phase, except for a lack of N in one out of every three atomic planes along the *c*-axis.[40, 42] $\eta$-Mn$_3$N$_2$ has a $T_N \sim 913$-927 K, and has the same spin ordering as $\theta$-MnN.[42]

*Magnetometry*

Magnetic hysteresis loops measured at 10 K and 300 K for the FC (black), FC+VC (red), and FC+VC-VC (blue) states are shown in Figs. 1b-c, respectively. The FC sample exhibits a large exchange field ($H_E$) of 1484 Oe at 10 K (Fig. 1b) and 618 Oe at 300 K (Fig. 1c). After positively biasing the samples (FC+VC state), $H_E$ increases from 1484 Oe to 1769 Oe ($\Delta H_E = 285$ Oe) at 10 K, a 19% increase, and from 618 to 646 Oe ($\Delta H_E = 28$ Oe) at 300K, a 5% increase. After a subsequent reverse biasing to the FC+VC-VC state, $H_E$ decreases in both cases (to 1741 Oe at 10 K and 631 Oe at 300 K). At 300K, $H_E$ of the FC state could essentially be recovered by negatively biasing for twice the duration (shown in Supporting Information Fig. S1). Following this recovery, applying another positive bias leads to an increase in $H_E$ again.

For a 1.0 nm thick Co$_{0.7}$Fe$_{0.3}$ film with a 0.25 cm$^2$ sample area, using the saturation magnetization value of 1510 $emu/cm^3$ [1 $emu/cm^3 = 1\ kA \cdot m^{-1}$], a saturation moment of $m_s =$ 38 μemu is expected. In the AG state, $m_s$ = 31 μemu, which is an ~18% reduction from the expected magnetic moment. Since the surface roughness of MnN, probed by X-ray reflectivity, is smaller than the $Co_{0.7}Fe_{0.3}$ thickness, this reduction in $m_s$ is not likely caused by any discontinuous FM layer. After field-cooling, a further $m_s$ decrease to 28 μemu is observed. Both decreases in $m_s$ are attributed to interfacial mixing at the MnN/$Co_{0.7}Fe_{0.3}$/Ta interfaces, observed in high angle annular dark field scanning transmission electron microscopy (HAADF-STEM) imaging, which is discussed in more detail later. However, no more substantial $m_s$ changes are seen for the FC+VC or FC+VC-VC states, where $m_s$ remains at 28 μemu. This fact suggests that the observed EB changes may be related to modifications in the MnN layer. To better understand the role the Ta capping layer plays in EB for this system, a reference sample of the same structure, but without the Ta capping layer, was measured in the FC state (shown in Supplemental Information Fig. S2). Under the same field cooling conditions, the reference sample shows almost no EB, indicating that the Ta capping layer plays a role in the initial EB, and may contribute to the enhancement seen in the FC+VC state.

*X-Ray Diffraction*

XRD measurements were used to investigate the film crystallinity in the AG, FC, and FC+VC states. The lattice parameters of MnN are known to sensitively depend on the N-content, e.g., the *c*-lattice constant increases with nitrogen content, with values ranging from 4.19 Å to 4.26 Å and beyond.[35, 43] To eliminate peak overlap between Pd (111), CoFe (200), and MnN (200) and (002), reference samples of $SiO_2$ /Ta (10 nm)/MnN (30 nm)/Ta (10 nm) were prepared using identical growth conditions. Grazing incidence XRD (GIXRD) confirms the *θ*-phase of MnN,[43] with (111), (200)/(002) and (220)/(202) peaks at 36.5°, 42.5°, and 61.7°, respectively (Fig. 2a). All extracted lattice parameters are in the range of 4.25 – 4.28 Å, with measurement uncertainty limiting the determinations of tetragonal distortions of the face-centered-cubic (fcc) lattice structure seen in $\theta$-MnN.[39, 40] The lattice parameters $c$ and $a$ in our system are equivalent from GIXRD, with $c/a \sim 1$. Previously, $c/a$ ratios of <1 are reported in $\theta$-MnN systems with N at% $\leq$ 50%,[39-41] and $c/a$ >1 are reported as N at% increases.[34-37] Thus, the $c/a$ ratio would indicate a N at% $\geq$ 50% in the MnN layer of our structure. Additional contributions from Ta are also observed, including tetragonal Ta (002) and (413) at 33.98° and 64.72°, respectively (PDF 00-025-1280), body-centered cubic (bcc) Ta (110) at 38.62° (PDF 00-004-0788), and a small $Ta_2O_5$ (220) peak at ~ 51.0° (PDF 00-018-1304).

Additionally, $\theta$-$2\theta$ scans were collected on a sample without the Ta capping layer, i.e. Ta (10 nm)/ MnN (30 nm)/ $Co_{0.7}Fe_{0.3}$ (1 nm)/ Pd (10 nm), as shown in Fig. 2b for the $2\theta = 40.0$-$45.5°$ range (full scan is shown in Supporting Information Fig. S2). This structure was chosen to avoid the observed Pd (111) peak when grown on the Ta capping layer, which overlaps with MnN (200)/(002). Instead, when grown directly on the 1 nm $Co_{0.7}Fe_{0.3}$ layer, the Pd (200) texture is promoted (PDF 00-046-1043). In the AG state (panel i), the (200)/(002) $\theta$-MnN peak is centered around 42.22°. This peak may be indexed as MnN (200) or (002) as no appreciable difference was seen between the (200) and (002) reflections from GIXRD. This corresponds to an out-of-plane lattice constant of 4.28 Å, which places the AG MnN in the upper limit of the $\theta$-MnN in terms of N-content.[40]

For the FC sample that has been exposed to 700 K while setting the EB, the MnN (200)/(002) peak exhibits a clear shift to 43.06°, along with an asymmetric shape (panel ii). This peak can be deconvoluted into two gaussian peaks at 42.91° and 43.52°, suggesting two MnN phases. The 42.91° peak corresponds to MnN with a lower N concentration compared to the AG state.[35, 39, 40] The latter peak at 43.52° is likely the (200)/(006) peak of the $\eta$ tetragonal phase of Mn$_3$N$_2$ with non-stoichiometric N-concentration, as their nominal positions are at 43.05° and 44.83°, respectively (PDF 00-001-1158). It is known that $\theta$-MnN first decomposes into $\eta$-Mn$_3$N$_2$ when annealed in vacuum, leading to a decrease in N-content.[43] Also, this $\eta$-phase shares the same crystalline space group as $\theta$-MnN (F4/mmm), where the only structural difference in the $\eta$-Mn$_3$N$_2$ is the lack of N in one out of every three atomic planes along the $c$-axis,[42] leading to a unit cell that is comparable to three $\theta$-MnN unit cells stacked along $c$. After voltage conditioning the sample, no appreciable further changes were found in the fitted peak positions or peak width (panel iii), each within the fitting error of the FC state. These findings will be further discussed in the electron microscopy and polarized neutron reflectometry sections below.

*Electron Microscopy*

High angle annular dark field scanning transmission electron microscopy (HAADF-STEM) imaging and STEM-EELS line-scan analysis were taken on Ta (10 nm) /MnN (15 nm)/ Co$_{0.7}$Fe$_{0.3}$ (1 nm)/ Ta (10 nm)/ Pd (10 nm) samples in the AG, FC, and FC+VC states. Fig. 3a shows a typical cross-sectional image of the AG multilayer stack. Analysis of the atomic images shows that $\theta$-MnN is crystallized in the AG state with a [001] texture (Fig. 3b-3c). In addition, a (Ta,Mn)N mixture is present at the bottom Ta/MnN interface of the stack. A mixture of Fe, Co, Mn, Ta, and N in the top Ta/CoFe interface is also observed, which accounts for the $m_s$ reduction observed from magnetometry discussed earlier. In the FC state, the MnN layer partially transformed to $\eta$-Mn$_3$N$_2$ which are predominantly out-of-plane ordered along [001] (Fig. 3d), suggesting that the deconvoluted peak in the XRD analysis is indeed the (200)/(006) reflections of the $\eta$-Mn$_3$N$_2$ phase. Fig. 3e displays an atomic image from a FC grain, which matches well with the overlapped atomic model of the $\eta$-Mn$_3$N$_2$ phase, lacking N in every third atomic layer, as compared with the θ-MnN in Fig. 3c. In the FC+VC state, $Mn_3N_2$ grains remained present in the MnN layer, and no significant crystalline changes were observed with HAADF-STEM.

STEM-EELS line-scan profiles, collected across the sample thickness, helped to identify changes in N-concentrations in each layer (Supporting Information Fig. S3). The elemental concentrations for Co, Fe, Mn, and N in the AG state reveal a higher relative N-concentration in the MnN layer compared to the Ta layers. In the FC state, this N distribution is altered, and the relative concentration in MnN decreases compared to the Ta layers. This is a manifestation of N moving out of MnN and into the Ta capping and seed layers. After gating the samples, no significant change in N concentration is observed with EELS.

These studies show that substantial changes occur in both the crystalline structure and N concentration in the MnN layer when the sample is field cooled from 700 K. In addition to the reduction of N-content in MnN seen by EELS, the formation of $Mn_3N_2$ grains does confirm that the net N-concentration in the MnN layer is decreasing, though the impact of these grains on EB is less clear. It is expected that $Mn_3N_2$ may not contribute strongly to the uniaxial anisotropy after field cooling despite being an AF, as field-cooling from 700 K is below its reported Neél temperature. Gating the sample did not produce changes that were observable with electron microscopy, indicating that the enhanced EB observed after gating is not due to significant structural changes in the heterostructure or from substantial N concentration changes at the level observable by EELS.

*Polarized Neutron Reflectometry*

Using PNR, the structural and magnetic depth profiles of the Ta (10 nm)/ MnN (30 nm)/ $Co_{0.7}Fe_{0.3}$ (1 nm)/Ta (10 nm)/ Pt (10nm) structure are probed in the AG, FC, and FC+VC states,[44-45] as shown in Fig. 4. An additional benefit of PNR for our structure is that small variations in N concentrations in the MnN layer should produce significant changes in the scattering length density (SLD). This is because Mn has a negative nuclear SLD ($\rho_N = -2.98 \times 10^{-4} nm^{-2}$), while MnN has a large positive nuclear SLD ($\rho_N = 1.77 \times 10^{-4} nm^{-2}$), leading to large contrast in $\rho_N$ between stoichiometric MnN and Mn. Fits of the PNR data for each state using the chosen model are shown in Supporting Information (Fig. S9), along with other excluded fitting models and a discussion of how the best model was chosen.

In the AG state (black curve in Fig. 4a), the MnN layer is best modeled with a continuous gradient in $\rho_N$. At the bottom Ta interface MnN is modeled with a $\rho_N$ of $2.88 \times 10^{-4} nm^{-2}$ (uncertainty for all $\rho_N < \pm 0.04 \times 10^{-4}$ nm$^{-2}$). At the top $Co_{0.7}Fe_{0.3}$ interface $\rho_N$ of MnN is modeled as $2.46 \times 10^{-4} nm^{-2}$. This range is somewhat higher than the nominal MnN $\rho_N$, indicating that the sample is N-rich, in agreement with the *c/a* ratio ~1 observed in XRD. The gradient may be due to variations in N$_2$ during sputtering or an intrinsic property of the MnN thickness, as observed in similar MnN heterostructures.[38] The $\rho_N$ for Ta and Pt in the AG state match closely with expected values, but the top region of the Ta seed and capping layers exhibit increased $\rho_N$. Deviation near the Ta capping layer is understandable, as the samples were exposed to air before depositing a Pt layer, indicating the formation of TaO$_x$ with a $\rho_N$ of $4.61 \times 10^{-4} nm^{-2}$ and a layer thickness of 4.5 nm, as stoichiometric Ta$_2$O$_5$ has a calculated $\rho_N$ of $4.79 \times 10^{-4} nm^{-2}$. The sharp increase in the Ta seed layer to $\rho_N = 7.17 \times 10^{-4} nm^{-2}$ at the Ta/MnN interface is likely due to exposure to N$_2$ plasma during deposition, leading to N inclusion in the layer. This value of $\rho_N$ is slightly above the nominal value of TaN ($\rho_N = 6.89 \times 10^{-4} nm^{-2}$), which may be an indication of an excess of N or intermixing of Mn into this region, as seen in HAADF-STEM. The magnetic component of the SLD, $\rho_M$, is indicated by the dashed black line in Fig. 4a. The only significant magnetic contribution is from the $Co_{0.7}Fe_{0.3}$ layer, though the model suggests a small magnetic inclusion in the top Ta layer, likely from Co, Fe, and Ta interdiffusion also observed by HAADF-STEM around the top $Co_{0.7}Fe_{0.3}$/Ta interface.

In the FC state (red curve in Fig. 4b), significant changes in layer thickness and $\rho_N$ are observed. In the MnN layer, $\rho_N$ decreases overall and is sufficiently modeled by two linear gradients decreasing from the center of the MnN layer. The values at the bottom, middle and top regions of the MnN layer are all below the nominal MnN $\rho_N$, with $\rho_{N,Bottom} = 1.21 \times 10^{-4} nm^{-2}$, $\rho_{N,Middle} = 1.69 \times 10^{-4} nm^{-2}$, $\rho_{N,Top} = 0.45 \times 10^{-4} nm^{-2}$, respectively. This depth profile indicates a significant reduction in N-concentration in this layer, in agreement with HAADF-STEM, EELS, and XRD, and has an associated reduction in MnN thickness of 1.3 nm. This model indicates the direction of N diffusion out of MnN is towards both the Ta seed and capping layer (including the $TaO_x$), as evidenced by the significant increase in $\rho_N$. Interestingly, the Ta seed layer seems to have a lower N concentration than the Ta capping layer, with $\rho_N = 4.91 \times 10^{-4} nm^{-2}$ and $\rho_N = 6.50 \times 10^{-4} nm^{-2}$, respectively. The cause of this asymmetric N diffusion is unclear, but it may indicate that the intermixed FM layer plays a role in catalyzing N diffusion in the system or that the TaN interface on the Ta seed layer may act as a N diffusion barrier.[36] As N diffuses into Ta, there also seems to be a significant increase in thickness of both Ta layers, with the seed layer increasing by 3.6 nm and the capping layer increasing by 0.8 nm, which may be explained by the lower density of TaN compared to Ta. $\rho_M$ in the FC state decreases in the $Co_{0.7}Fe_{0.3}$ layer from $5.89 \times 10^{-4} nm^{-2}$ in the AG state to $4.54 \times 10^{-4} nm^{-2}$, which is consistent with the observed decrease in $m_s$.

In the FC+VC state, PNR provides insight into the effect of applying an electric field and the corresponding increase in $H_E$ (Fig. 4b blue curve). First, a decrease in $\rho_N$ is observed in the Ta seed layer ($\rho_N = 4.82 \times 10^{-4} nm^{-2}$) and in the bottom half of the MnN layer ($\rho_{N,Bottom} =$ $0.87 \times 10^{-4} nm^{-2}$; $\rho_{N,Middle} = 1.62 \times 10^{-4} nm^{-2}$), which corresponds to a decrease in N-concentration. An increase in $\rho_N$ relative to the FC state is also observed in the top half of the MnN layer ($\rho_{N,Top} = 0.67 \times 10^{-4} nm^{-2}$) and the Ta capping layer ($\rho_N = 6.73 \times 10^{-4} nm^{-2}$), indicating the net motion of N under bias is +Z, toward the top contact. The change in $\rho_N$ in the MnN and Ta layers is statistically significant, as the 95% confidence intervals of the modeled $\rho_N$ do not have any overlap in the bottom of MnN ($95\% CI_{FC} = 1.19 - 1.25 \times 10^{-4} nm^{-2}$; $95\% CI_{FC+V} = 0.85 - 0.90 \times 10^{-4} nm^{-2}$), the top portion of MnN ($95\% CI_{FC} = 0.41 - 0.49 \times 10^{-4} nm^{-2}$; $95\% CI_{FC+VC} = 0.64 - 0.70 \times 10^{-4} nm^{-2}$), the Ta seed layer ($95\% CI_{FC} = 4.88 - 4.94 \times 10^{-4} nm^{-2}$; $95\% CI_{FC} = 4.79 - 4.85 \times 10^{-4} nm^{-2}$), or the Ta capping layer ($95\% CI_{FC} = 6.46 - 6.54 \times 10^{-4} nm^{-2}$; $95\% CI_{FC+VC} = 95\% CI = 6.70 - 6.77 \times 10^{-4} nm^{-2}$). Small changes in layer thicknesses (0.2 - 0.4 nm for Ta and MnN layers), leading to the total thickness offset in Fig. 4b, and insignificant changes in magnetization from the FC to FC+VC were also seen in this model.

**Discussion**

The interesting electric-field enhancement of the EB can be understood by nitrogen ionic migration. In the AG state, N is present in the top region of the Ta seed layer, which may occur due to spontaneous gettering of N by Ta from the MnN layer. Using the stoichiometries for tabulated thermodynamic properties of Ta and $\theta$-phase MnN ($Mn_6N_5$), the reaction would follow Eqn. 1 below. The spontaneous gettering of N from MnN by Ta is supported by the calculated Gibbs free energy of -132.6 kJ/mol and a calculated enthalpy of formation of -137.3 kJ/mol,[46-48] indicating the interfacial reaction would be both spontaneous and exothermic.

$$5Ta + Mn_6N_5 \rightarrow 6Mn + 5TaN \quad (1)$$

Furthermore, during the FC process, N diffusion is enhanced by the elevated temperatures and N moves to both the Ta capping layer and seed layers since the direction of the reaction remains the same, as seen in EELS and PNR. This subsequently leads to both induced EB while cooling the sample in a magnetic field and a reduction in N content in the MnN layer. STEM and magnetometry both suggest that intermixing at the MnN/Co$_{0.7}$Fe$_{0.3}$/Ta interfaces occur in the AG state, and exposure to elevated temperatures during FC leads to further intermixing.

After the FC+VC process, PNR's depth resolution, particularly its sensitivity to N, indicates that N indeed migrates in the structure. The direction of this motion is +Z under a positive bias, with N moving toward the top of the MnN layer, as well as into the Ta capping layer. No difference in the amount of Mn$_3$N$_2$ is seen by HAADF-STEM or XRD after gating, suggesting this N increase is within MnN itself. The increase in $H_E$ can be attributed to this N motion, as N content increases in MnN near the Co$_{0.7}$Fe$_{0.3}$/MnN interface after gating, which is supported by literature where increased exchange bias with N-content is well observed in MnN systems due to increases in the interfacial exchange constant, $J_{ex}$, with increasing N-content.[35-36, 38-39] This effect may be considered as equivalently an increase in the effective AF layer thickness, consistent with a 0.2 nm increase in the MnN layer thickness observed in PNR, that helps to provide a stronger pinning of the FM, thus a larger EB. Another possible source of $H_E$ enhancement can be attributed to the observed N diffusion into the Ta capping layer under biasing. It is likely that defects are introduced at the top MnN/Co$_{0.7}$Fe$_{0.3}$ interface as a result of the nitrogen migration, leading to changes in pinned uncompensated AF moments, which sensitively influences $H_E$.[2, 4, 49-52] This is highlighted by the observation of significantly different EB values for this structure compared to the reference sample without a Ta capping layer (Fig. S2). Though it could only be demonstrated for the FC state since the reference sample had essentially no EB, it does indicate that N moving to the Ta capping layer during the field cooling process is an important component of the observed EB. Driving more N to this Ta layer in the FC+VC states could then also play a significant role in the EB enhancement. Even a small amount of N-migration may cause sufficient modification of interface to result in a significant change in EB. No significant contributions to EB enhancement are expected to be caused by $Mn_3N_2$ in the MnN layer, as the FC process occurred well below the $T_N$ of $\eta$-$Mn_3N_2$ and thus is not expected to alter the interfacial FM/AF coupling.

In the FC+VC-VC state, the bias is reversed across the sample and N migration is expected to be from the Ta capping layer and upper half of MnN toward the lower half of MnN and the Ta seed layer. The decrease in EB under negative biasing, along with an increase in EB following a subsequent positive biasing indicates the potential for reversible control of EB with electric fields. While the data shown in Fig. S1 is representative of the continued gating trend in most samples of similar structure, it should be noted that the first cycle (positive and negative bias) trend is consistent across all samples and that deviations have been seen in a small number of samples after the first cycle. These deviations in trend are believed to be a result of the sample structure, with two relatively symmetric N reservoirs (Ta layers partially converted to TaN) around MnN. These layers provide a source of N regardless of the gating direction, which will be addressed by future studies.

The requirement of longer gating times to achieve full reversibility under negative biasing can be understood from the thermodynamic properties of the system. First, the negative Gibbs energy of the reaction in Eqn. 1 indicates that the formation of TaN is preferred, which would require less energy to drive the reaction forward. Conversely, the energy required to reverse the direction of N migration will be greater due to this same factor. Additionally, even as MnN decomposes to $Mn_3N_2$ as seen in HAADF-STEM, the calculated Gibbs free energy is -212.6 kJ/mol for the reaction shown in Eqn. 2,[47-48] indicating even as N is lost, the remaining Ta will preferentially form TaN over the reverse reaction.

$$2Ta + Mn_3N_2 \rightarrow 3Mn + 2TaN \qquad (2)$$

**Conclusions**

In summary, significant nitrogen-based magneto-ionic enhancement of exchange bias has been observed in Ta/MnN/Co$_{0.7}$Fe$_{0.3}$/Ta heterostructures, which can be electrically manipulated. A comprehensive set of studies using magnetometry, HAADF-STEM, EELS, and PNR has enabled probing of structural, magnetic, and N-concentration changes across the structure under different sample conditions. When samples are field cooled, a clear N migration out of the MnN layer and into the Ta layers is evident, and the formation of Mn$_3$N$_2$ grains occur. This field-cooling step leads to a significant exchange bias. Upon positive voltage biasing, $H_E$ increases by 19% at 10K and 5% at room temperature. This enhancement corresponds to both N migration in the +Z direction into the top half of the AF MnN layer, as well as into the Ta capping layer. It is both the increased N content in MnN and likely changes to the pinned uncompensated AF spins at the interface that contribute to the enhancement of EB. Reverse biasing for a longer duration leads to the recovery of the initial EB, with a subsequent positive bias indicating this electric control of EB may persist beyond one cycle. These results demonstrate the potential for electrical manipulation of exchange bias via the magneto-ionic handle. The modulation achieved using the nitrogen ions has potential applications in low energy-dissipation nanoelectronics.

**Methods**

**Sample synthesis.** Thin films of Ta seed (10 nm)/MnN (30 nm)/Co$_{0.7}$Fe$_{0.3}$ (1 nm)/Ta (10 nm)/Pd or Pt cap (10nm) (bottom to top structure) were grown on thermally oxidized *p*-type Si substrates (SiO$_2$ = 285 nm) following a standard cleaning procedure in acetone, isopropanol, and deionized water. A shadow mask was used to pattern the samples into 5 mm × 5 mm arrays. All sputtering depositions were performed in a chamber with a base pressure of <6 × $10^{-6}$ Pa, and a working pressure of 0.33 Pa. The 10nm Ta seed layer was first grown by direct current (DC) magnetron sputtering. The 30 nm MnN was then RF reactively sputtered from an elemental Mn target with a 1:1 $N_2$:Ar mixture.[35, 39] The 1 nm CoFe was grown onto MnN by either DC co-sputtering from elemental Co and Fe targets or from a single $Co_{0.7}Fe_{0.3}$ composite target. In the co-sputtering case, deposition power was calibrated to achieve a 70:30 ratio of Co:Fe. Finally, the samples were capped with 10 nm Ta, removed from the chamber (<1 h air exposure), and reintroduced to the chamber for a 10 nm Pd or Pt top electrode deposition. A reference sample without a Ta capping layer for magnetometry comparison was also prepared with the same conditions (10 nm)/MnN (30 nm)/$Co_{0.7}Fe_{0.3}$ (1 nm)/Ta (10 nm)/Pd or Pt cap (10nm).

**Magnetic measurements.** Magnetic characterizations were conducted using a superconducting quantum interference device (SQUID) magnetometer (Quantum Design MPMS3). To establish exchange bias, the sample was heated inside the SQUID to 700 K at a rate of 50 K/min, above the Néel temperature of MnN ($T_N \sim 655 - 660\ K$).[41-42] Once this temperature was reached, a 6.5 kOe in-plane magnetic field was applied and held for 1 minute. Subsequently, the sample was cooled to 300 K or 10 K in this field at a cooling rate of 50 K/min. Hysteresis loops were measured with a saturation magnetic field of 20 kOe for samples in the as-grown (AG), field cooled (FC), and subsequent voltage conditioned (VC) states (FC+VC and FC+VC-VC). Here, FC+VC refers to a positive gating of 20 V for 1 hour after field cooling, using a Keithley 2280S Precision Measurement DC supply, which was also used to monitor any potential oxide breakdown. The FC+VC-VC refers to the subsequent negative biasing for 1 hour after the initial positive gating. After field cooling each sample, magnetic field treatments were used to decrease the field training effect observed in the samples to below the measurement error (details in Supporting Information). The reference sample without the Ta capping layer was also field cooled and measured under the same conditions at 300K for comparison.

**X-ray diffraction.** Structural characterization by X-ray diffraction was performed using a Malvern-Panalytical X'Pert3 MRD system with Cu $K_\alpha$ radiation in both $\theta$-$2\theta$ and grazing incidence (GIXRD) configurations. The $\theta$-$2\theta$ scans were performed over a $2\theta$ range of 20˚-130˚ using a PixCel line detector with a step size of 0.02˚ and total integration time of 1000 s. GIXRD scans used an incidence angle of 0.5˚, a Xe proportional detector (point detector) with a step size of 0.05˚ and a total integration time of 60 s over a range of 30˚-70˚ in $2\theta$.

**Neutron scattering.** Polarized neutron reflectometry (PNR) measurements were carried out at NIST Center for Neutron Research on the Polarized Beam Reflectometer. Measurements were taken at room temperature with a 15 kOe magnetic field applied in plane along the field-cooling axis of the samples. The neutron beam was polarized parallel (+) or antiparallel (-) to the magnetic field, and non-spin-flip specular reflectivities ($R_{++}$ and $R_{--}$) were measured with respect to wave vector transfer, Q. The REDUCTUS and Refl1D software packages were used to reduce and fit the data, respectively.[44-45] Error bars were determined with a Markov chain Monte Carlo method using the BUMPS software package.

**Electron microscopy.** Electron microscopy studies were performed at the NIST Materials Measurement Laboratory. Electron transparent cross-sectional samples were prepared with an FEI Nova NanoLab 600 DualBeam (SEM/FIB). An FEI Titan 80-300 probe-corrected STEM/TEM microscope operating at 300 keV was employed to conduct atomic-resolution high-angle annular dark field scanning transmission electron microscopy (HAADF-STEM) imaging and electron energy-loss spectroscopy (EELS) analysis.

**Supporting Information:** The Supporting Information is available free of charge online at (URL). Additional experimental details, methods, and analysis relating to the gating response, magnetometry, and PNR (PDF).

**Acknowledgement**

This work has been supported in part by SMART (2018-NE-2861), one of seven centers of nCORE, a Semiconductor Research Corporation program, sponsored by National Institute of Standards and Technology (NIST), by the NSF (ECCS-2151809), and by KAUST (OSR-2019-CRG8-4081). The acquisition of a Magnetic Property Measurements System (MPMS3), which was used in this investigation was supported by the NSF-MRI program (DMR-1828420). H.Z. acknowledges support from the U.S. Department of Commerce, NIST under financial assistance award 70NANB19H138 and 70NANB22H101. A.V.D. acknowledges support from the Material Genome Initiative funding allocated to NIST. Disclaimer: Certain commercial equipment, instruments, software, or materials are identified in this paper in order to specify the experimental procedure adequately. Such identifications are not intended to imply recommendation or endorsement by NIST, nor it is intended to imply that the materials or equipment identified are necessarily the best available for the purpose.

**Figures**

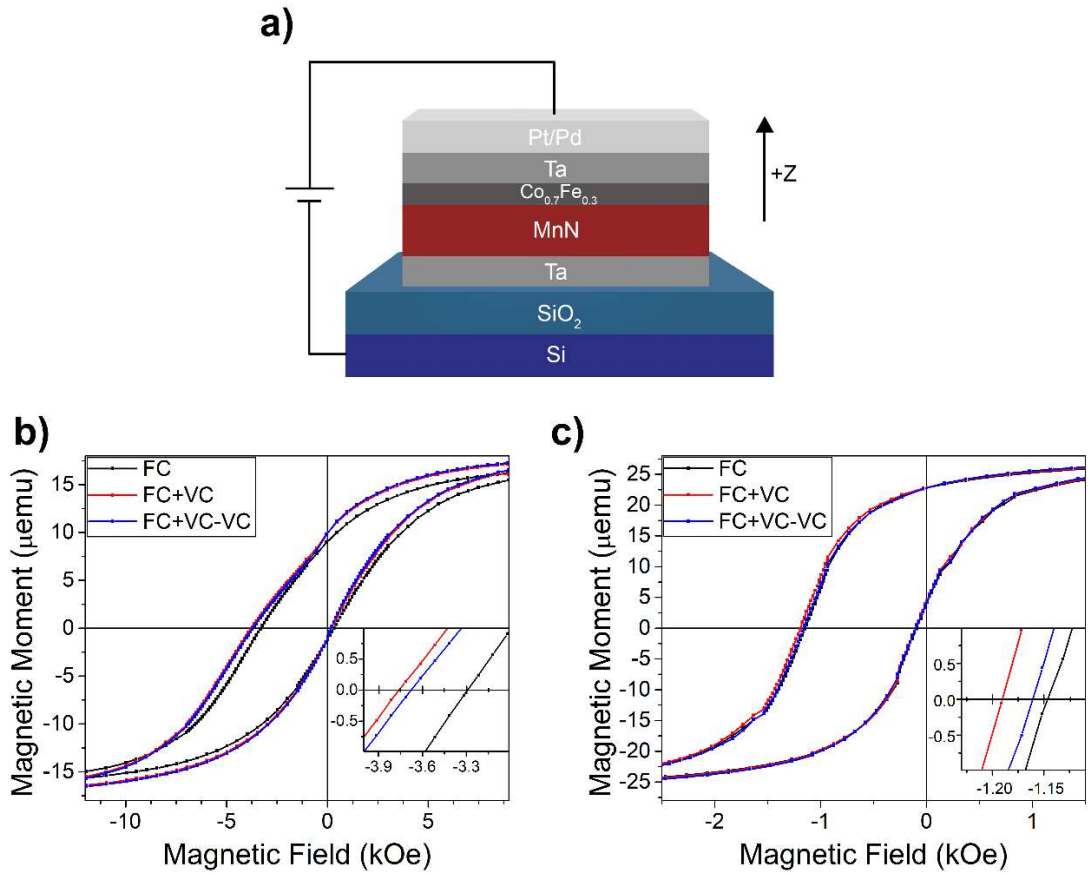

**Figure 1**. a) Schematic diagram of the Ta/MnN/Co$_{0.7}$Fe$_{0.3}$/Ta/Pd(Pt) heterostructure showing positive gating convention and arrow indicating the spatial direction considered positive (+Z) in the heterostructure. Hysteresis loops are shown of a Ta (10 nm)/ MnN (30 nm)/ Co$_{0.7}$Fe$_{0.3}$ (1 nm)/Ta (10 nm)/ Pd (10nm) sample at b) 10K and c) 300K. Each measurement was taken after preparing the same sample in the FC (black), FC+VC (red), and FC+VC-VC (blue) states. Insets show a zoomed-in view of the descending branch where the total magnetic moment passes through 0.

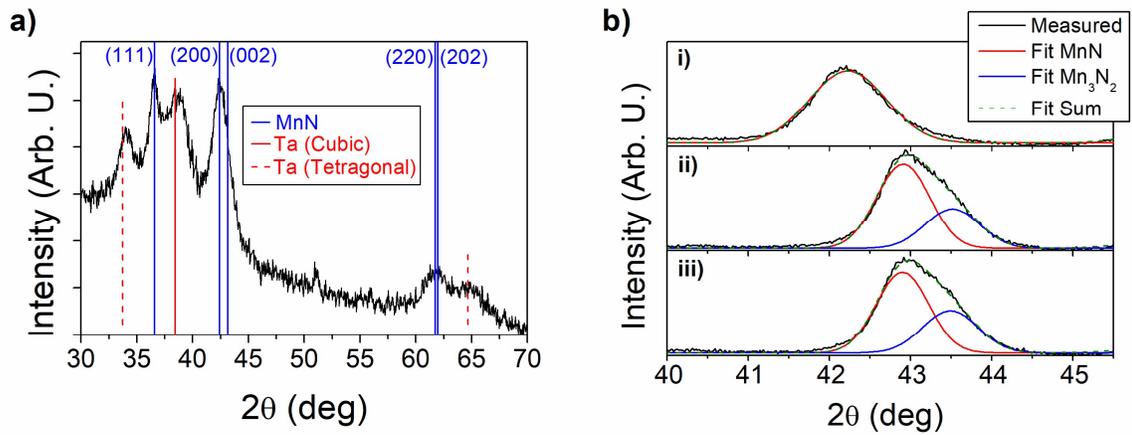

**Figure 2**. a) Grazing incidence X-ray diffraction (GIXRD) of Ta (10 nm)/MnN (30 nm)/Ta (10 nm). Blue solid lines correspond to tabulated peaks for the θ-phase of MnN,[43] solid red lines for the cubic phase of Ta, and dashed red lines for the tetragonal phase of Ta. b) X-ray diffraction $\theta$-$2\theta$ scan of Ta (10 nm)/MnN (30 nm)/CoFe (1nm)/Pd (10 nm) in the i) AG, ii) FC, and iii) FC+VC states. The black line is the measured data, the red and blue line represents the gaussian fit for the peak associated with MnN and $Mn_3N_2$, respectively, and the green dashed line shows the sum of the two fitting curves. i) is only fit with one gaussian, corresponding to MnN.

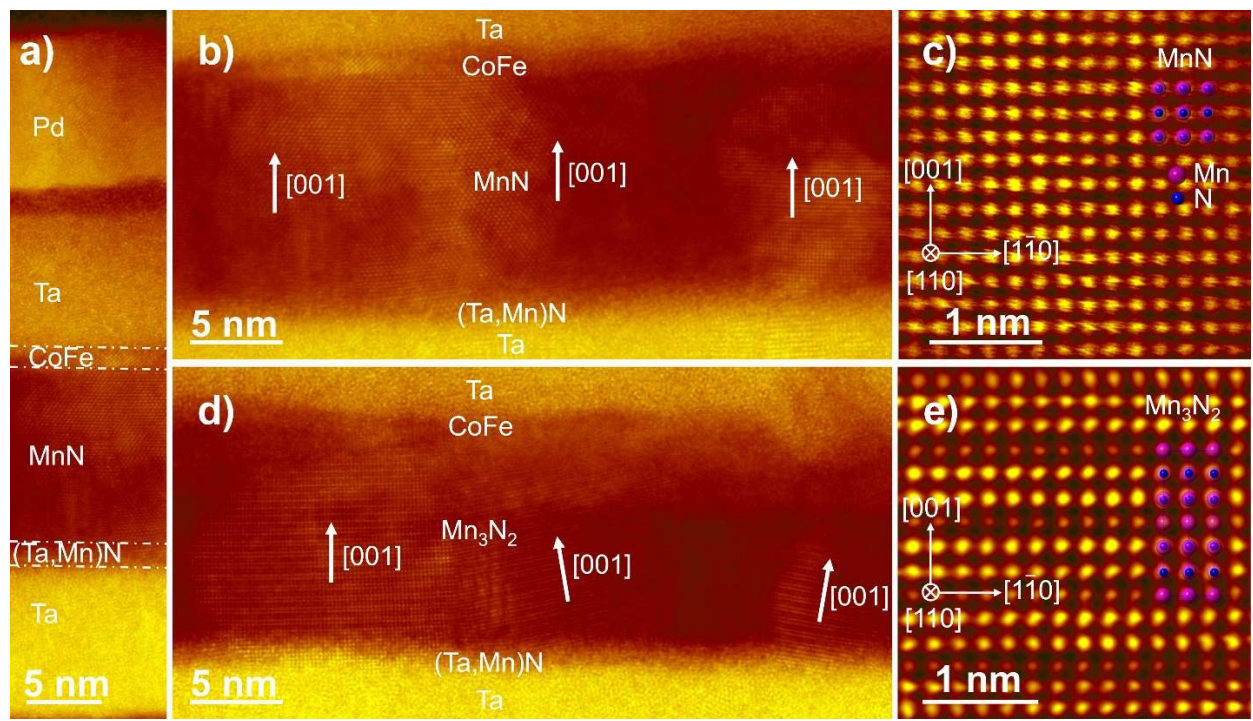

**Figure 3**. HAADF-STEM images of Ta (10 nm)/MnN (15 nm)/Co$_{.7}$Fe$_{0.3}$ (1 nm)/Ta (10 nm)/Pd (10 nm) for a) the full structure, b-c) AG state, and d-e) FC state. a) has dashed lines indicating the interface between Ta/MnN at the bottom and MnN/Co$_{0.7}$Fe$_{0.3}$/Ta at the top. For the AG state, b) arrows indicate the [001] texture of MnN, and c) shows the [010] zone-axis atomic structure of MnN. For the FC state d) arrows indicate the [001] texture of Mn$_3$N$_2$ and e) shows the [010] zone-axis atomic structure of Mn$_3$N$_2$.

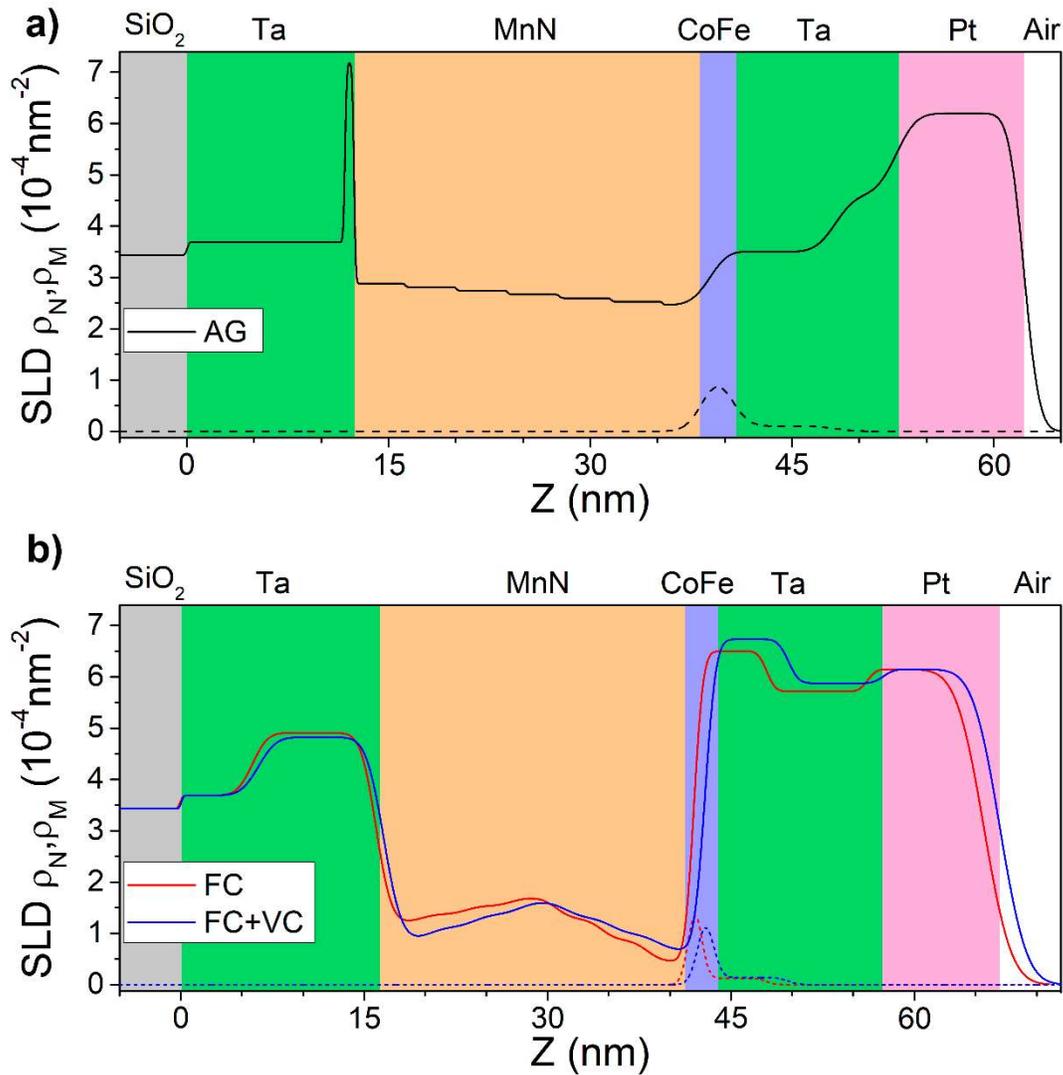

**Figure 4.** SLD depth profiles for a) the AG state (black) and b) the FC (red) and FC+VC (blue) states of Ta (10 nm)/MnN (30 nm)/Co$_{.7}$Fe$_{0.3}$ (1 nm)/Ta (10 nm)/Pt (10 nm). Solid lines represent the nuclear component of the SLD, and dashed lines represent the magnetic component of the SLD. The background colors indicate locations of various layers in the sample heterostructure.

Supporting Information

# Nitrogen-Based Magneto-Ionic Manipulation of Exchange Bias in CoFe/MnN Heterostructures

Christopher J. Jensen[1], Alberto Quintana[1], Patrick Quarterman[2], Alexander J. Grutter[2], Purnima P. Balakrishnan[2], Huairuo Zhang[3,4], Albert V. Davydov[4], Xixiang Zhang,[5] and Kai Liu[1,*]

[1]Physics Department, Georgetown University, Washington, DC 20057, USA
[2]NIST Center for Neutron Research, NCNR, National Institute of Standards and Technology, Gaithersburg, MD 20899, USA
[3]Theiss Research, Inc. La Jolla, California 92037, USA
[4]NIST Materials Measurement Laboratory, National Institute of Standards and Technology, Gaithersburg, MD 20899, USA
[5]King Abdullah University of Science & Technology, Thuwal 23955-6900, Saudi Arabia

**Gating Modulation of Exchange Bias**

Repeated voltage gating was performed on the sample shown in the main text magnetometry result (Fig. 1). The sample was first subjected to a positive bias for 1 hour, which led to the enhancement of exchange bias, as described in the main text. Fig. S1 shows the exchange field, $H_E$, for this as well as subsequent gating of -20V for 1 hour, a repeated -20V for 1 hour, and a final +20V for 5 hours. $H_E$ is approximately revered after the second negative biasing, reaching a value of 620 Oe, and positively gating for 5 hours led to an increase of $H_E$ to 628 Oe.

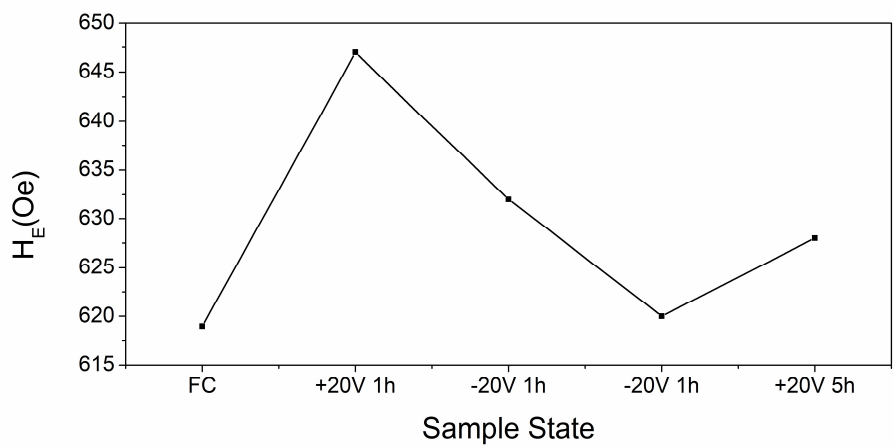

**Figure S1**. Exchange bias gating trends of the Ta (10 nm)/MnN (30 nm)/CoFe (1 nm)/Ta (10 nm)/Pd (10 nm) sample. The *x*-axis denotes the gating/preparation step performed before the corresponding $H_E$ value was measured, with each step being sequentially performed on the same sample.

Supporting Information

**Capping Layer Dependence of Exchange Bias**

A reference sample was grown of the structure in the main text, but without a Ta capping layer before the Pd contact was deposited. Remarkable differences in $H_E$ was observed, as shown in Fig. S2, with the reference sample exhibiting almost no EB. In addition, there is little magnetic coercivity seen. This indicated that Ta plays a significant role in establishing EB for the conditions used.

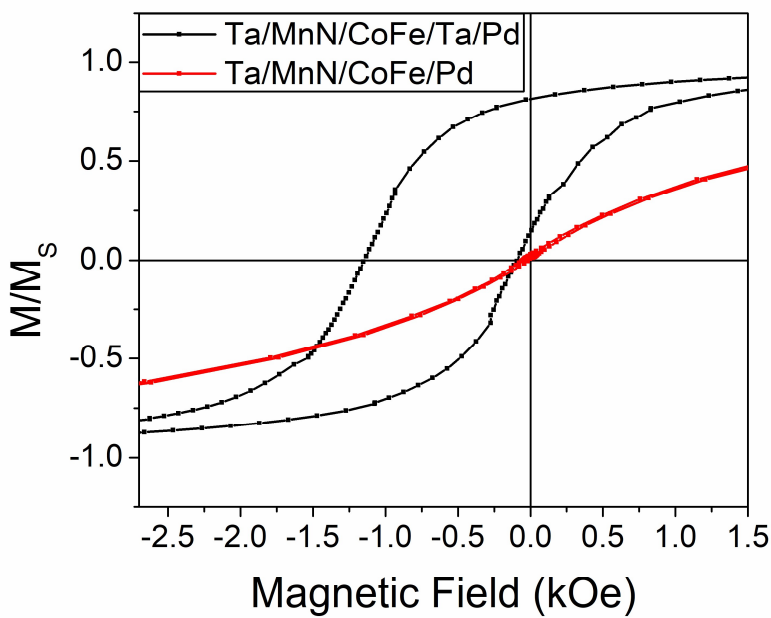

**Figure S2**. The FC state hysteresis loops at 300 K of the Ta (10 nm)/MnN (30 nm)/CoFe (1 nm)/Ta (10 nm)/Pd (10 nm) structure compared to a reference sample without a Ta capping layer of Ta (10 nm)/MnN (30 nm)/CoFe (1 nm)/Pd (10 nm). The magnetic moment is normalized to the saturation moment of each sample.

Supporting Information

**X-Ray Diffraction**

X-ray diffraction $\theta$-$2\theta$ scans over an extended range of 20°-110° in $2\theta$ are shown in Figure S3 for each state of the Ta (10 nm)/ MnN (30 nm)/ Co$_{0.7}$Fe$_{0.3}$ (1 nm)/ Pd (10 nm) heterostructure. The same scans are presented in the main text in Fig. 2b over the $2\theta$ range of 40.0°-45.5° to highlight changes in the MnN layer. In addition, a reference scan on an identical thermally oxidized Si substrate is included, measured with the same beam configuration as the samples.

The MnN (002)/(200) and Mn$_3$N$_2$ (006)/(200) peak in Fig. 2b are discussed in the main text of this paper, but to confirm (002)/(200) texture in MnN all other peaks must be identified. Peaks for each scan in the 46°-80° region can all be attributed to the substrate and background of the instrument, as each peak present is also in the substrate reference scan. Additional peaks were identified for the AG, FC, and FC+VC states at 37.3°, 40.3°, and 104.9°. Tabulated peaks for tetragonal Ta have peaks corresponding to the (330) reflection at 37.4° and for the (411) reflection at 40.2° (PDF 00-025-1280). The 104.9° peak corresponds to the (400) Pd tabulated peak at 104.8° (PDF 00-046-1043). An additional peak is present in the FC and FC+VC states at 93.4°, which corresponds to the (004)/(400) reflections of $\theta$-phase MnN at 93.5°[1] and the (400) reflection of $\eta$-phase Mn$_3$N$_2$ at 94.4° (PDF 00-001-1158). One final peak is present in the FC+VC state, located at 33.0°, which due to the sharpness of the peak is attributed to a forbidden reflection of Si. The (002)/(200) texture of MnN in the AG state, as well as the c-axis texture corresponding to the convoluted peaks of (002)/(200) MnN and (006)/(200) Mn$_3$N$_2$ in the FC and FC+VC states are supported by no other MnN or Mn$_3$N$_2$ reflections being present in the θ-2θ scan data.

Supporting Information

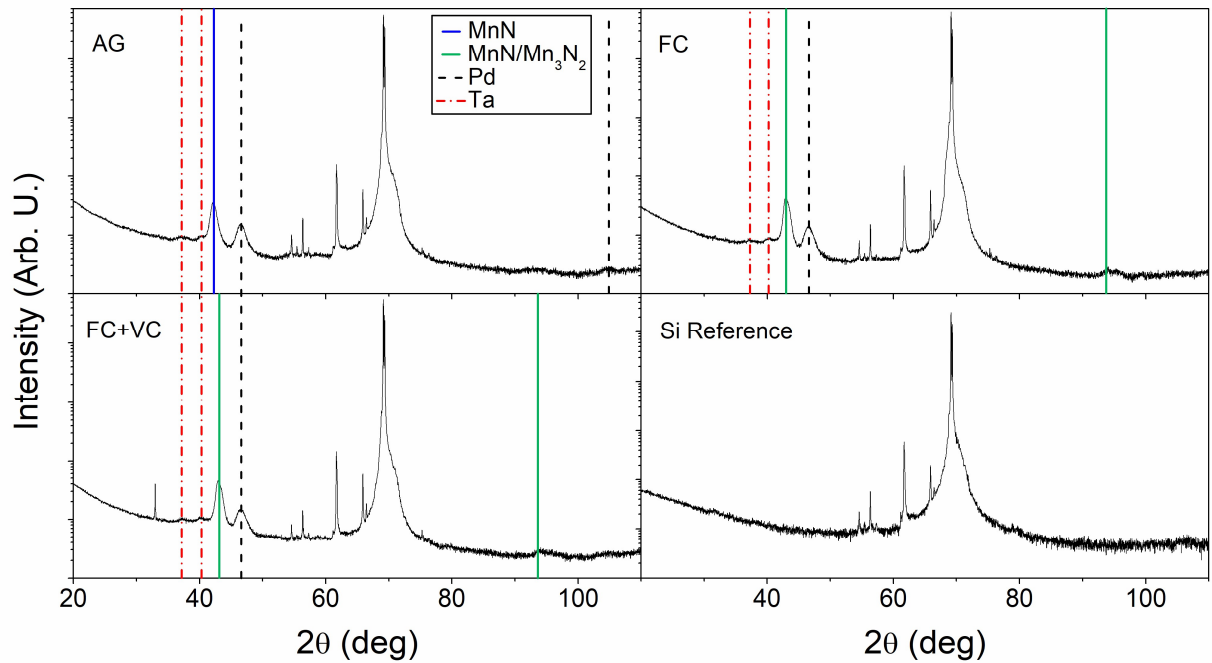

**Figure S3**. Full range X-ray diffraction $\theta$-$2\theta$ scans of the AG, FC, and FC+VC states of Ta (10 nm)/MnN (30 nm)/CoFe (1 nm)/Pd (10 nm) sample and a reference Si substrate. Each plot is on a log scale in intensity and has MnN, MnN/Mn$_3$N$_2$, Pd, and Ta peaks labeled. Unlabeled peaks correspond to Si and detector background reflections.

**STEM-EELS**

STEM-EELS line-scans profiles were taken for cross-sections of the Ta/MnN/Co$_{0.7}$Fe$_{0.3}$/Ta/Pd structure in the AG, FC, and FC+VC states. The AG state (Fig. S4a) shows a higher relative concentration of N in the MnN layer compared to both Ta layers. After the structure is exposed to 700 K during the field cooling process (Fig. S4b), the relative concentration of N in the MnN layer and Ta layers are approximately equal. This suggests that N moves from the MnN layer and into both the Ta seed and capping layer.

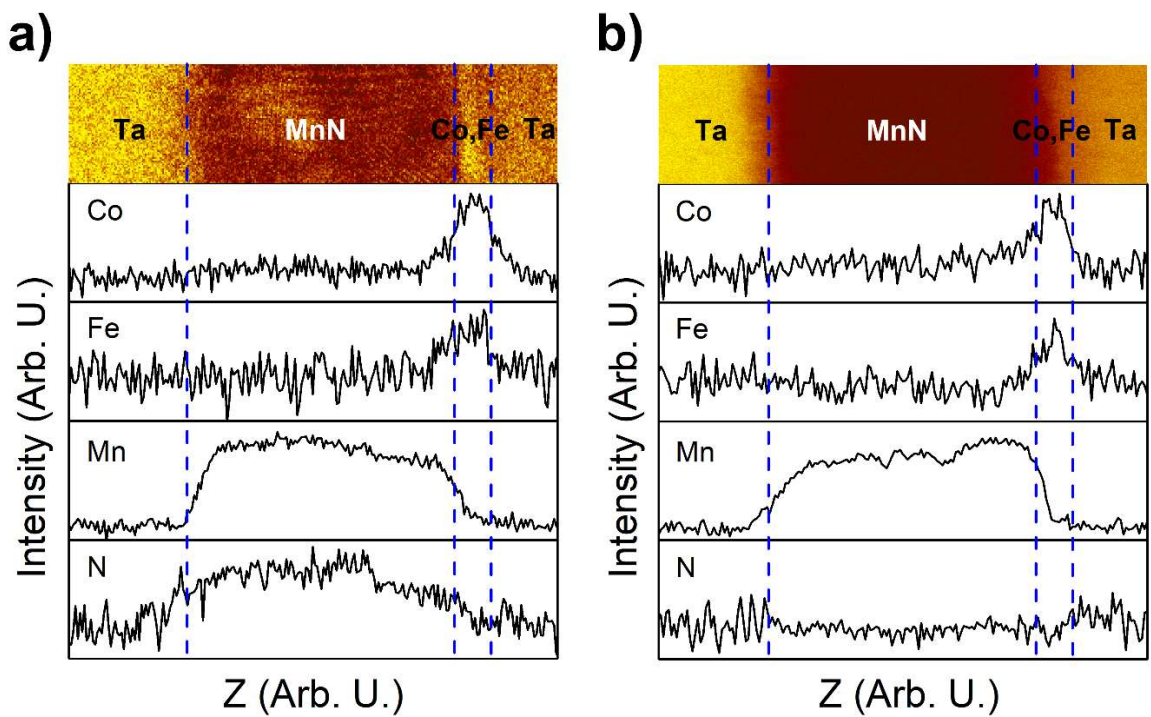

**Figure S4**. STEM-EELS line-scan profiles for the Ta (10 nm)/MnN (15 nm)/Co$_{.7}$Fe$_{0.3}$ (1 nm)/Ta (10 nm)/Pd (10 nm) structure in the a) AG and b) FC states. The top panels show cross sections of the HAADF-STEM images, and each panel is labeled with the representative element. Dashed black lines are used to highlight the interfaces.

**Polarized Neutron Reflectometry Modeling**

   Polarized Neutron Reflectometry (PNR) measurements were conducted at NIST Center for Neutron Research. Reduction and fitting of the data were performed in the REDUCTUS and Refl1D software packages.[2-3] While these packages offer powerful fitting algorithms, model optimization remains a challenge when performing simultaneous refinement off multiple datasets with linked models and a wide parameter space, as is often the case for magneto-ionic systems. Similarly, solution degeneracy means that more than one model is likely to sufficiently describe a reflectivity dataset. Thus, models must be chosen such that the fitting is not over or under parameterized while enforcing physically reasonable boundary conditions. To achieve this, we chose to start with the simplest models and work toward more complex models, ultimately choosing the best model that was supported by additional measurements on the heterostructure, such as electron microscopy. For each model, the AG, FC, and FC+VC states were fit simultaneously in order to enforce parameters uniformity for layers, such as Si and $SiO_2$, across different sample states that were expected to remain unchanged. By fitting in this way, we avoid unphysical solutions in which layers not participating in voltage-driven electrochemical reactions change between field conditions, while simultaneously achieving improved statistical certainty on the derived parameters. In this section, we will describe the models that were excluded and how we arrived at our chosen model presented in the main text of the paper.

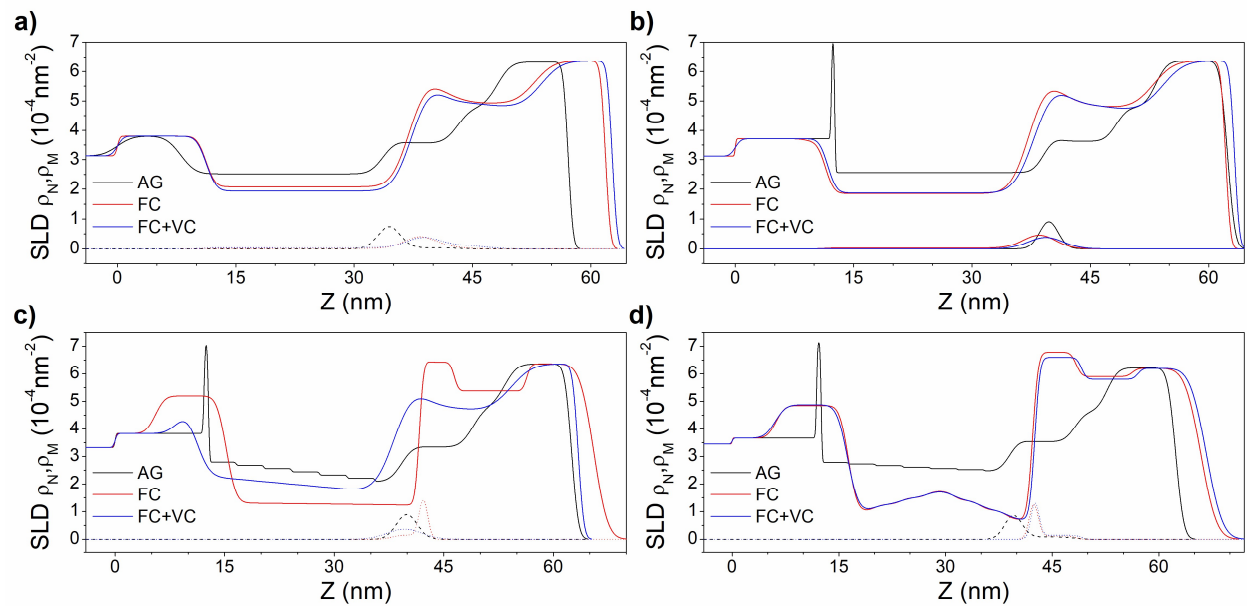

**Figure S5**. SLD depth profiles for the AG state (black), FC state (red), and FC+VC state (blue) of Ta (10 nm)/MnN (30 nm)/Co$_{.7}$Fe$_{0.3}$ (1 nm)/Ta (10 nm)/Pt (10 nm). The panels correspond to a model using a) constant nuclear components of the SLD, b) constant nuclear component of the SLD and inclusion of interfacial N in the Ta seed layer, c) linear gradients in the nuclear component of the SLD, and d) linear gradients in the nuclear components of the SLD in the MnN layer that were allowed to vary in two different regions.

## *Constant SLD Model*

For the simplest model, each layer in the heterostructure was fit using a constant nuclear, $\rho_N$, and/or magnetic, $\rho_M$, component of the scattering length density, SLD. The layers included in the model were Ta, MnN, Co$_{0.7}$Fe$_{0.3}$, Ta, TaO$_x$, and Pt from bottom to top of the structure, respectively. TaO$_x$ was included due to the sample being removed from the deposition chamber between the Ta and Pt growth, which likely caused oxidation. Layer thicknesses and roughness were given initial values estimated from growth parameters, and $\rho_N$ and $\rho_M$ were calculated from density and saturation magnetization for each layer.

The SLD profile for this model is shown in Fig. S5a, where a variation in thickness from the AG state to FC and FC+VC states is indicated. Qualitative analysis of the fittings, shown in Fig. S6, can be used to immediately see that this model does not fit the data well between 0.7 to 1.2 nm$^{-1}$ in Q for the AG state (Fig. S6a). Additionally, poor fitting is seen for both the FC (Fig. S6c) and FC+VC (Fig. S6e) conditioned states. Quantitatively, this model has the highest $\chi^2$ for the AG ($\chi^2$ = 3.67) and FC state ($\chi^2$ = 3.55) of any model. Spin asymmetry [$SA = (R_{++} - R_{--})/(R_{++} + R_{--})$] is calculated for both the measured data and simulated fits and plotted for the AG (Fig S6b), FC (Fig. S6d), and FC+VC (Fig. S6f) states. SA arises from reflectivity differences in the $R_{++}$ and $R_{--}$ reflectivity cross sections when interacting with magnetic materials, but also contains structural information. The poor fits in the 0.7 to 1.2 nm$^{-1}$ range of Q for SA additionally highlight the ineffectiveness of this model. This model was excluded for these reasons.

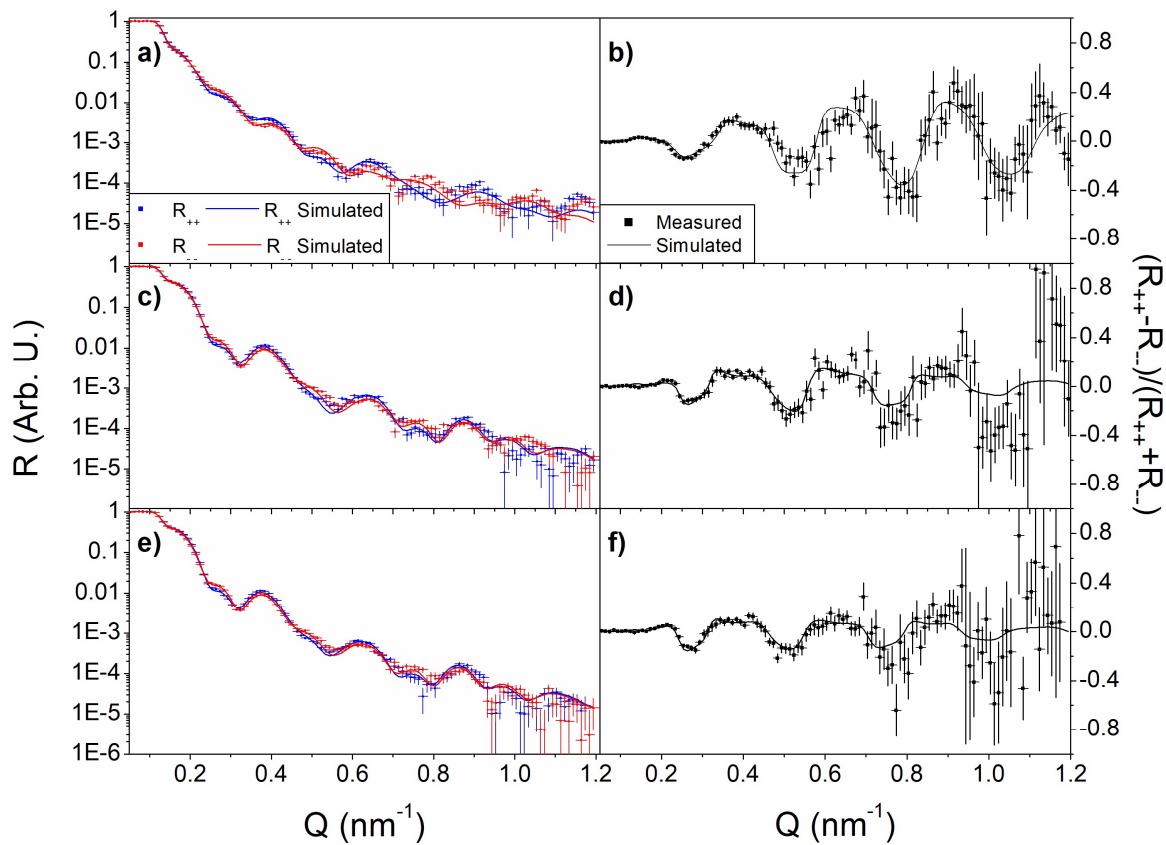

**Figure S6**. a,c,e) Fits of the reflectivity data and b, d, f) calculated spin asymmetry using the sample model of constant nuclear SLD values in each layer of Ta (10 nm)/MnN (30 nm)/Co$_{.70}$Fe$_{0.3}$ (1 nm)/Ta (10 nm)/Pt (10 nm). Reflectivity fits and spin asymmetry calculations are presented for the a,b) AG, c,d) FC and e,f) FC+VC states, respectively.

### *Constant SLD + TaN Model*

To improve on the previous model, a TaN layer was added to the model at the Ta/MnN interface. This layer was observed in the PNR analysis of similar samples by Quarterman et al.[1] and is expected to form due to N$_2$ exposure to Ta during sputter growth of MnN. This layer was again fit with a constant $\rho_N$. All other parameters follow with the previous model. Fig. S5b shows the modeled SLD profile corresponding to this model, with the only significant change seen in the AG state, where a sharp peak with increase $\rho_N$ is present at the Ta/MnN interface, corresponding to TaN ($\rho_N(Ta) = 3.82 \times 10^{-4} nm^{-2}$; $\rho_N(TaN) = 7.19 \times 10^{-4} nm^{-2}$).

Supporting Information

From the fitted PNR data, shown in Fig. S7, an improvement is seen in the AG state fit (Fig. S7a), especially in the 0.7 to 1.0 nm$^{-1}$ range in Q, which is also reflected in the SA seen in Fig. S7b. Additionally, $\chi^2$ slightly improves for the AG ($\chi^2$ = 3.50) and FC state ($\chi^2$ = 3.35), though the fits for the FC (Fig. S7c) and FC+VC (Fig. S7e) states qualitatively do not match the data well. SA for the FC (Fig. S7d) and FC+VC (Fig. S7f) show little to no improvement in the 0.7 to 1.2 nm$^{-1}$ range in Q, and the SA fit is qualitatively worse for the FC+VC state, especially around 0.3 nm$^{-1}$. Because of the poor fits for the FC and FC+VC states, this model was excluded.

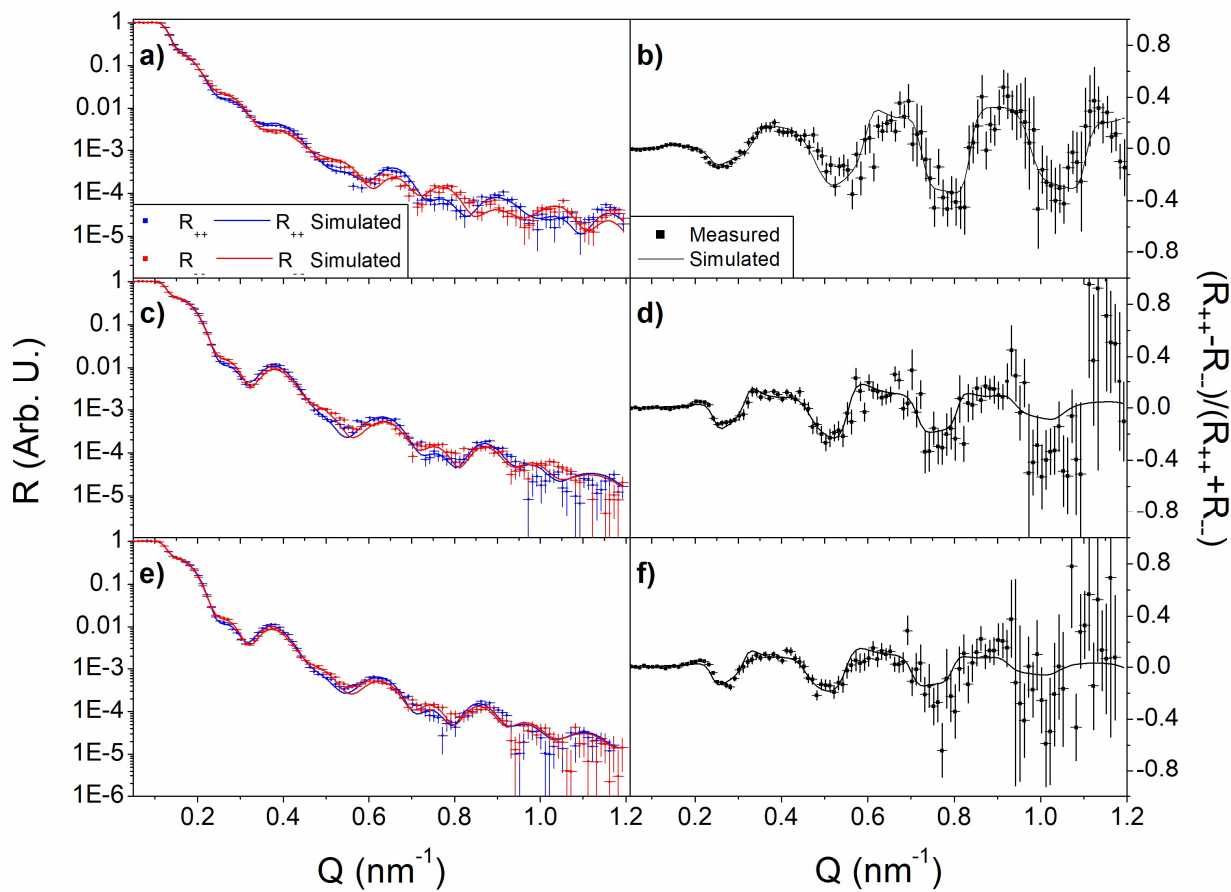

**Figure S7**. a,c,e) Fits of the reflectivity data and b, d, f) calculated spin asymmetry using the sample model of constant nuclear SLD values in each layer of Ta (10 nm)/MnN (30 nm)/Co$_{.7}$Fe$_{0.3}$ (1 nm)/Ta (10 nm)/Pt (10 nm) and allowing interfacial N at the Ta/MnN seed layer interface. Reflectivity fits and spin asymmetry calculations are presented for the a,b) AG, c,d) FC and e,f) FC+VC states, respectively.

Supporting Information

*MnN Linear Gradient Model*

As it became evident that constant $\rho_N$ values were not sufficient to model the PNR data, the next logical choice was to allow linear gradients to be fit across layers where there may be structural and/or magnetic variations. To avoid introducing too many parameters at once, a linear gradient was only added to the MnN layer in the structure since N variations within this layer may be significant, especially between the AG to FC state as seen in the STEM-EELS measurements shown in Fig. S4. To achieve this gradient, MnN was broken into 7 separate layers, with the top-most and bottom-most layers being fit. MnN layers 2-6 were calculated to create a linear gradient with equal step sizes in $\rho_N$ from the top to bottom layers. This linear gradient was allowed to vary such that $\rho_N$ on the top region of MnN could be higher, lower, or equal to the bottom region. All other parameters in this model were kept the same as the previous model, still allowing a TaN layer to be present.

Fig. S5c shows the fit SLD profiles for each state of the sample for this model. A decreasing linear gradient in $\rho_N$ from the bottom to the top of the MnN layer is present in the AG and FC+VC states, while the FC state has a relatively constant $\rho_N$. TaN again seems to be present at the Ta/MnN interface in the AG state, indicated by the sharp increase in $\rho_N$, and a TaN layer is also present in the fits for the FC and FC+VC conditioned states. Additional interface roughness changes are seen in the SLD profiles, where the roughness increase across most multilayer interfaces from the FC to FC+VC states. Large structural changes such as these are not expected following voltage biasing, so this model does not fit the physical picture of the sample well. Additionally, the fit PNR data shown in Fig. S8 has improvements qualitatively and in $\chi^2$ for the AG (Fig. S8a; $\chi^2$ = 2.64) and FC state (Fig. S8c; $\chi^2$ = 3.13), but the fit for the FC+VC (Fig. S8e) state worsens. A similar trend is also seen in the SA where the FC state (Fig. S8d) has a closer match between

Supporting Information

simulated model and experimental data at higher Q values, and the FC+VC state (Fig. S8f) worsens overall. Here, we note that because of the linked parameters across all three models, the $\chi^2$ variation of the fit to any given layer is not always intuitive. Rather, the software optimizes the overall integrated $\chi^2$ for the linked model across all datasets. Thus, while a given model might fit one condition better, the overall fit across all conditions must be considered when evaluating which models to use. Because of the poor fit in the FC+VC state and non-physical interface roughness increases from the FC to FC+VC states, this model was excluded.

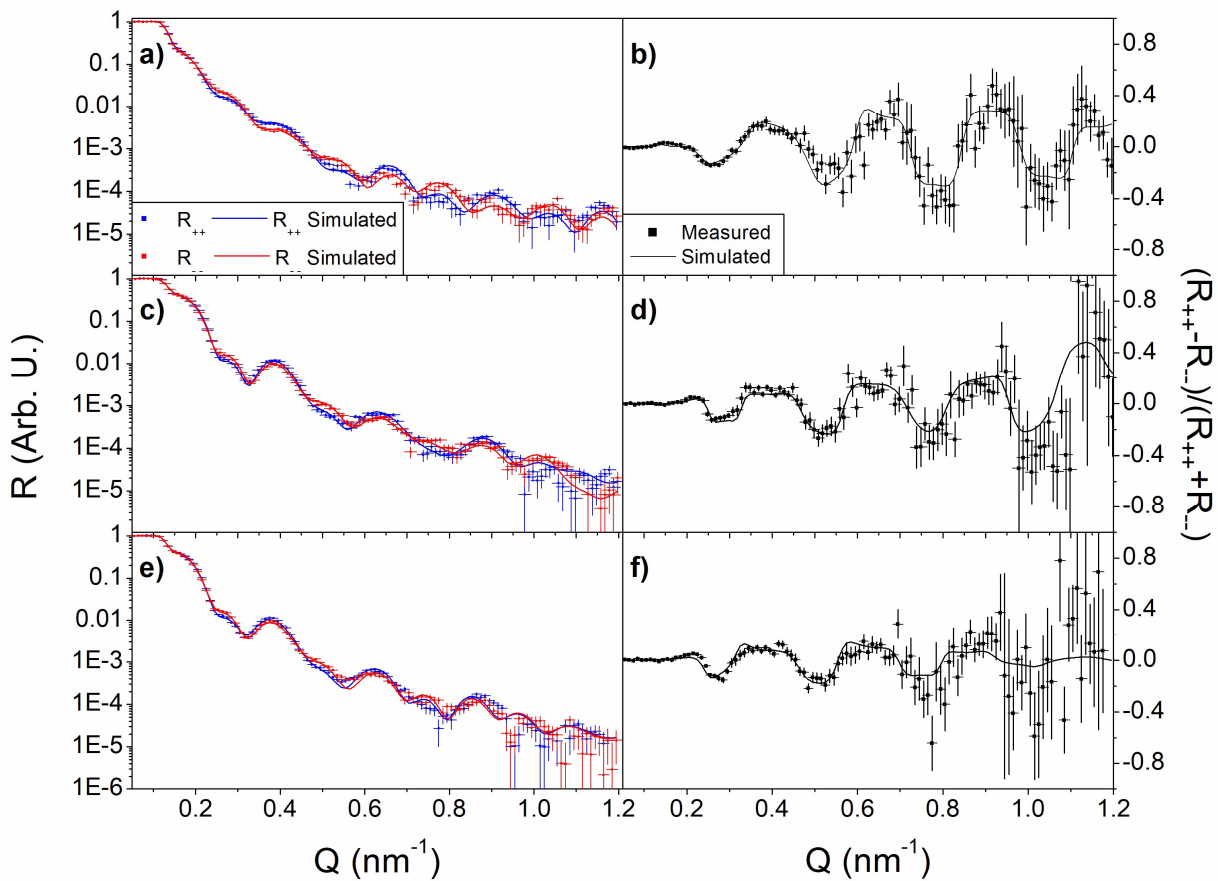

**Figure S8**. a,c,e) Fits of the reflectivity data and b,d,f) calculated spin asymmetry using the sample model of linear gradients in the MnN layer for the nuclear SLD values in of Ta (10 nm)/MnN (30 nm)/Co.$_7$Fe$_{0.3}$ (1 nm)/Ta (10 nm)/Pt (10 nm). Reflectivity fits and spin asymmetry calculations are presented for the a,b) AG, c,d) FC and e,f) FC+VC states, respectively.

*Chosen Model - Multiple Linear Gradient*

The chosen model for the PNR data came from the next step in modeling variation. For this, instead of a constant linear gradient across MnN, two separate linear gradients were allowed to be fit. This was chosen due to the Ta capping and seed layers, which are N getters, likely pulling N from both sides of the MnN layer. To achieve the gradients, the MnN layer was broken into the same 7 layers, but with the top-most, bottom-most, and middle layer being fit. Layers 2, 3, 5, and 6 are calculated to create equal size steps in $\rho_N$ between the bottom to middle layer and the middle to top layer. All other layer parameters are the same as the previous model.

A detailed discussion of the fit SLD profiles for this model are presented in the main text and shown in Fig. 4. PNR data fits for each state are shown in Fig. S9. Significant qualitative and $\chi^2$ improvements are seen for the AG (Fig. S9a; $\chi^2 = 1.74$), FC (Fig. S9c; $\chi^2 = 1.42$), and FC+VC state (Fig. S9e; $\chi^2 = 1.35$). Additionally, SA significantly improves in the 0.7-1.2 nm$^{-1}$ range in Q for the FC+VC state (Fig. S9f), while the simulated fits for the AG (Fig. S9b) and FC (Fig. S9d) states still match the overall data SA profile well. Because of the good qualitative fits (especially in the 0.7-1.2 nm$^{-1}$ range in Q), consistancy in physical meaning of $\rho_N$ and $\rho_M$ across the layers for each state, and best $\chi^2$ values for any model used, this model was chosen as the most representative.

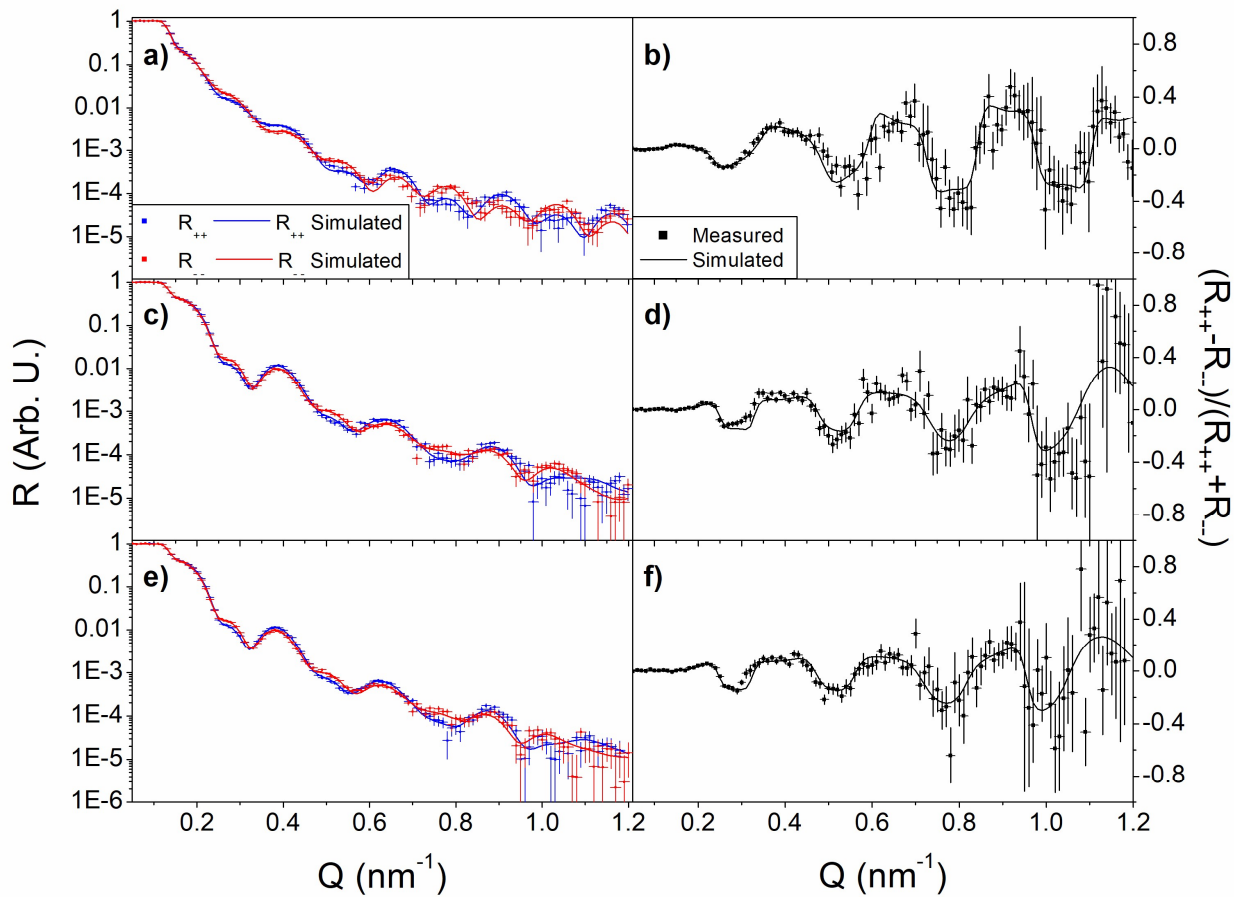

**Figure S9**. a,c,e) Fits of the reflectivity data and b,d,f) calculated spin asymmetry using the sample model of linear gradients in the MnN layer for the nuclear SLD values, with separate gradients able to exist in the top and bottom half of the layer in Ta (10 nm)/MnN (30 nm)/Co$_{.7}$Fe$_{0.3}$ (1 nm)/Ta (10 nm)/Pt (10 nm). Nuclear scattering length densities in the MnN layer for the FC and FC+VC states are set to be equal. Reflectivity fits and spin asymmetry calculations are presented for the a,b) AG, c,d) FC and e,f) FC+VC states, respectively.

*Multiple Linear Gradient + Linked Sample State Model*

The last model, with fit SLD profiles shown in Fig. S5d, was chosen to help determine if the $\rho_N$ variation seen from the FC to FC+VC state in the previous model are significant. To do this, the exact parameters from the previous model were used, but with the $\rho_N$ for the top, middle, and bottom regions of MnN forced to be equal between the FC and FC+VC states. If there was no significant difference in $\rho_N$ between the two states, they should be fit with similar $\chi^2$ values and qualitatively match when using an identical MnN layer.

Supporting Information

Fig. S10 shows the fit PNR data for this model. $\chi^2$ for each state increases, with the largest increase in the FC ($\chi^2 = 1.50$) and FC+VC state ($\chi^2 = 1.43$). While the differences in $\chi^2$ and qualitative data fitting are not large, this model was excluded since it does not provide as good of a fit as the previous model. Additionally, this provides evidence that the $\rho_N$ difference from the FC to FC+VC state are small but significant, with further evidence provided from non-overlapping 95% confidence intervals in MnN for the chosen model, as discussed in the main text.

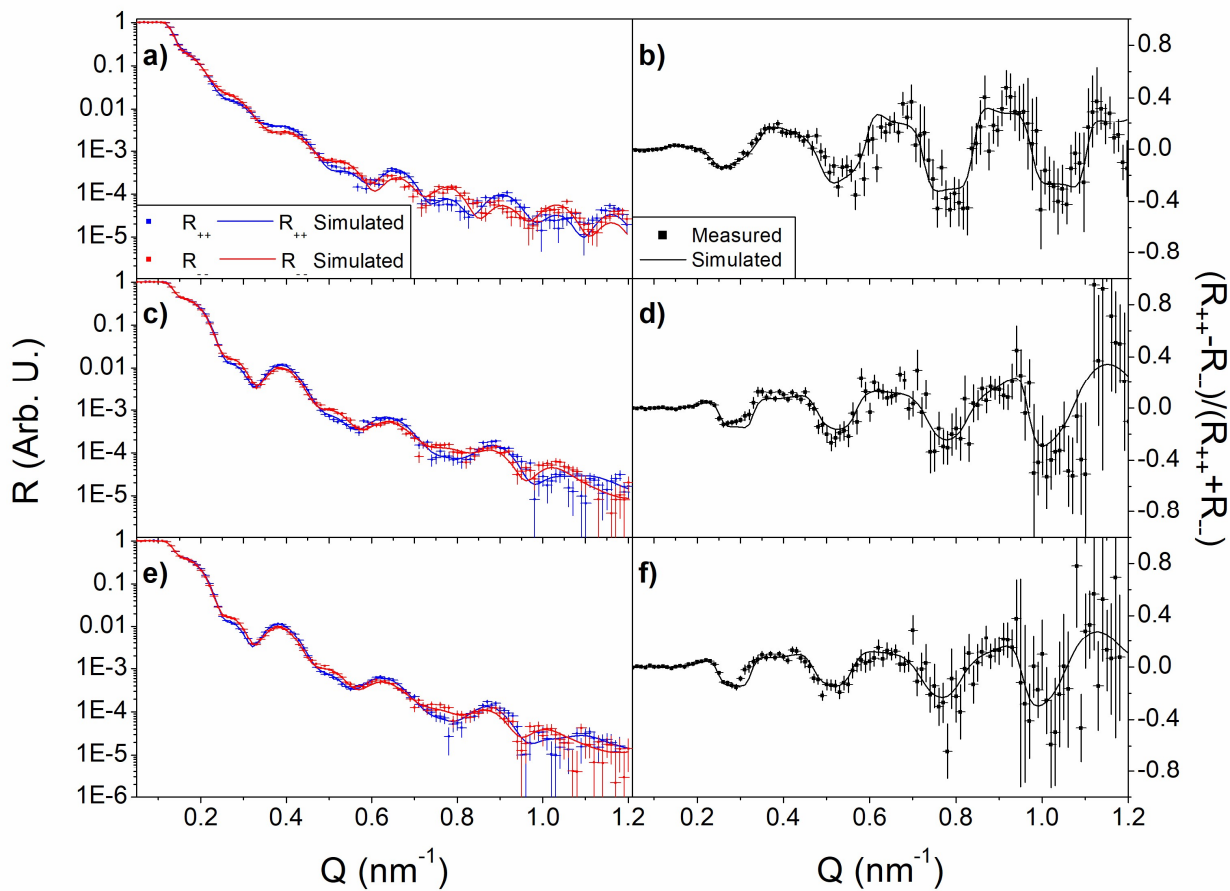

**Figure S10**. a,c,e) Fits of the reflectivity data and b,d,f) calculated spin asymmetry using the sample model of linear gradients in the MnN layer for the nuclear SLD values, with separate gradients able to exhist in the top and bottom half of the layer in Ta (10 nm)/MnN (30 nm)/Co$_{.7}$Fe$_{0.3}$ (1 nm)/Ta (10 nm)/Pt (10 nm). Reflectivity fits and spin asymmetry calculations are presented for the a,b) AG, c,d) FC and e,f) FC+VC states, respectively.

Supporting Information

**Field Training**

To remove the field training effect often associated with exchange bias systems, after field cooling the samples at room temperature a 70 kOe field was applied for 5 minutes, followed by 10 cycles of 20 kOe to -20 kOe and holding the field at -20 kOe for an additional 30 minutes, which reduced the field training effect below measurement error. Exchange bias field, $H_E$, and coercive field, $H_C$, measured during repeated cycles after applying a 70 kOe field for 5 minutes are shown in Fig. S11. After 10 cycles, $H_C$ has stabilized, and changes in $H_E$ are almost an order of magnitude smaller ($\Delta H_E$ = -6 Oe) than after the first cycle ($\Delta H_E$ = -42 Oe). Holding the samples at -20 kOe for 30 minutes following cycling further decreases changes in repeated $H_E$ measurements.

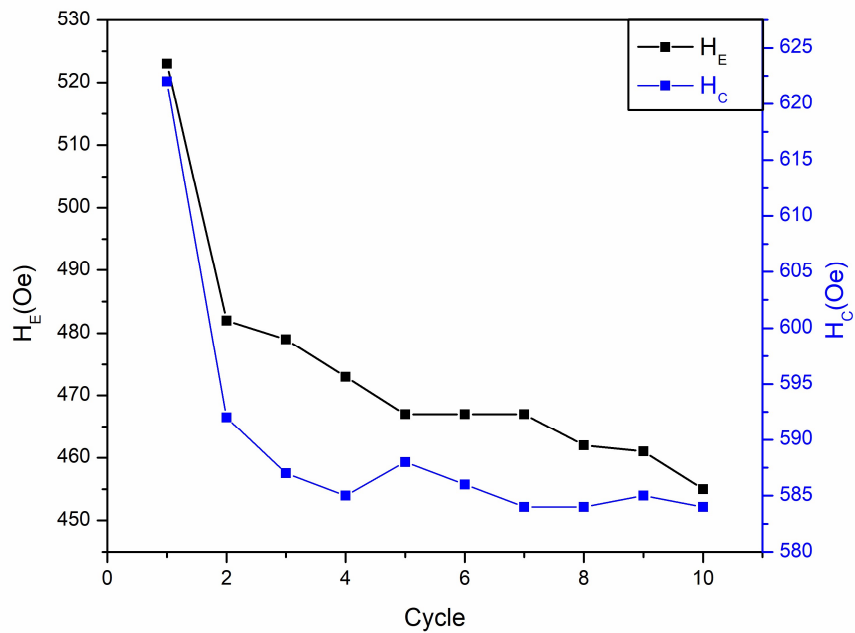

**Figure S11**. Exchange bias field, $H_E$, and coercive field, $H_C$, for the Ta (10 nm)/MnN (30 nm)/Co$_{.7}$Fe$_{0.3}$ (1 nm)/Ta (10 nm)/Pt (10 nm) measured repeatedly after applying a 7 T field for 5 minutes at room temperature.

Supporting Information